\documentclass [11pt]{article}

\usepackage{slashbox}
\usepackage{graphicx}
\usepackage{amsmath,amssymb} 
\usepackage{latexsym}
\usepackage{mathrsfs}
\usepackage{bbold}
\usepackage{color}
\usepackage{slashed}
\definecolor{red}{rgb}{1,0,0}
\usepackage{cancel}
\makeatletter
\def\section{\@startsection {section}{1}{\z@}{-3.5ex plus -1ex minus
 -.2ex}{2.3ex plus .2ex}{\large\bf}}
\def\subsection{\@startsection{subsection}{2}{\z@}{-3.25ex plus -1ex
minus -.2ex}{1.5ex plus .2ex}{\normalsize\bf}}
\makeatother
\makeatletter

\@addtoreset{equation}{section}

\makeatother

\def\bea{\begin{eqnarray}} \def\eea{\end{eqnarray}}
\def\be{\begin{equation}} \def\ee{\end{equation}} \def\nn{\nonumber}

\newcommand{\SO}{\text{SO}}
\newcommand{\SU}{\text{SU}}
\newcommand{\U}{\text{U}}

\newcommand{\promille}{%
  \relax\ifmmode\promillezeichen
        \else\leavevmode\(\mathsurround=0pt\promillezeichen\)\fi}
\newcommand{\promillezeichen}{%
  \kern-.05em%
  \raise.5ex\hbox{\the\scriptfont0 0}%
  \kern-.15em/\kern-.15em%
  \lower.25ex\hbox{\the\scriptfont0 00}}

\usepackage[dvips,colorlinks,linkcolor=black,citecolor=blue,urlcolor=blue,linktocpage]{hyperref}
\newcommand{\hhref}[2][]{\href{http://arxiv.org/abs/#2#1}{arXiv:#2}}

\begin{document}

\thispagestyle{empty}

\begin{center}

\hfill SISSA-32/2012/EP \\

\begin{center}

\vspace*{0.5cm}

{\Large\bf  UV Completions of Composite Higgs Models} \\ [3mm]
{\Large\bf  with Partial Compositeness}

\end{center}

\vspace{1.4cm}

{\bf Francesco Caracciolo$^{a}$, Alberto Parolini$^{a}$ and Marco
Serone$^{a,b}$}\\

\vspace{1.2cm}

${}^a\!\!$
{\em SISSA and INFN, Via Bonomea 265, I-34136 Trieste, Italy} 

\vspace{.3cm}

${}^b\!\!$
{\em ICTP, Strada Costiera 11, I-34151 Trieste, Italy}

\end{center}

\vspace{0.8cm}

\centerline{\bf Abstract}
\vspace{2 mm}

\begin{quote}
We construct UV completions of bottom-up 
models with a pseudo Nambu-Goldstone Boson (NGB) composite Higgs and partial compositeness, admitting 
a weakly coupled description of the composite sector.
This is identified as the low energy description of an $\SO (N)$ supersymmetric gauge theory
with matter fields in the fundamental of the group. The Higgs is a NGB associated to an $\SO(5)/ \SO(4)$ coset of a global symmetry group
and is identified with certain components of matter fields in a Seiberg dual description of the theory. The Standard Model (SM) gauge fields are obtained by gauging
a subgroup of the global group. The mass mixing between elementary SM and composite fermion fields  advocated in partial compositeness 
arise from the flow in the IR of certain trilinear Yukawa couplings  defined in the UV theory. 
We explicitly construct two models of this kind.
Most qualitative properties of the bottom-up constructions are derived. The masses of gauge and fermion resonances in the composite sector
are governed by different couplings and can naturally be separated. Accommodating all SM fermion masses within the partial compositeness paradigm remains the main open problem, since
the SM gauge couplings develop Landau poles at unacceptably low energies.
\end{quote}

\vfill

\newpage

\tableofcontents

\section{Introduction}

A possible solution to the gauge hierarchy problem is to assume that the Higgs field is a  pseudo Nambu-Goldstone boson (pNGB) of a spontaneously broken approximate global symmetry 
in a strongly coupled theory. A light Higgs at about 125 GeV, parametrically lighter than the scale where the resonances of the strong sector arise,  is elegantly explained by its
NG nature, very much like pions in QCD. Despite this idea is quite old  \cite{Kaplan}, considerable progress has been obtained in the framework of five-dimensional (5D) theories, where
possible viable models have been proposed \cite{ACP,GHU1}.  
The 5D picture revealed that the most successful Composite Higgs Models with a pNGB Higgs (from now on denoted pCHM) feature another important property,
called partial compositeness (again, an old idea  \cite{Kaplan:1991dc}, whose consequences have been fully exploited only thanks to 5D model building \cite{Grossman:1999ra,Contino:2006nn}).
Such pCHM contain essentially two sectors, an ``elementary'' sector, including the SM gauge and fermion fields,  and
a ``composite'' strongly coupled sector, including the Higgs field (and possibly the right-handed top quark $t_R$) and heavy resonances. The global symmetry is explicitly broken
by gauging a subgroup of it via the SM gauge interactions and by quadratic terms which mix the SM
fermions with fermion resonances of the  strong sector. Due to these mixing, SM vectors and fermions become partially composite. In particular, 
the lighter are the SM fermions, the weaker are the mixing. This simple, yet remarkable, observation allows to significantly alleviate most flavor bounds.
With these insights, a purely 4D model building featuring a pNGB Higgs and partial compositeness is possible, where the composite sector is
(relatively) weakly coupled and can be described in terms of free fields, see e.g.\cite{Giudice:2007fh,Gripaios:2009pe,Contino:2011np,Marzocca:2012zn}.

While, on the phenomenological side,  the attention should now be devoted to the study of  the LHC signatures of pCHM, a fundamental theoretical problem is still open.
As far as we are aware, no UV completion of pCHM with partial compositeness has been proposed so far.\footnote{A UV model with a composite pNGB Higgs has been constructed in \cite{Galloway:2010bp}, but with fully elementary SM fermions and no partial compositeness.}

The aim of this paper is to look for possible UV completions of pCHM.  Most likely one needs a full theory of gravity, such as string theory, 
to UV complete the 5D pCHM.  Finding non-supersymmetric string vacua resembling even at a rough qualitative level the 5D pCHM  seems a formidable task.
For this reason we focus our attention on UV completions of the 4D bottom-up pCHM, where the composite sector in the IR is weakly coupled and might admit a completion in terms of some quantum field theory.

Following the RG flow of a strongly coupled theory is a hard task. We circumvent this problem by assuming that the composite sector is approximately supersymmetric (SUSY). 
Supersymmetry is also helpful in explaining the appearance of relatively light, meson-like, fermion resonances,
necessary to implement partial compositeness for the SM fermions.\footnote{A scenario with a composite Higgs in an approximately SUSY strong sector has  been considered in \cite{Redi:2010yv}.} Without fundamental scalars, fermion bound states are expected to be baryon-like and at higher scales.
For concreteness, we consider in this paper completions of the minimal pCHM based on the $\SO(5)/\SO(4)$ coset structure, although generalization to other cosets should be straightforward.
We take as candidate UV completions of the composite sector ${\cal N}=1$ SUSY non-abelian gauge theories with gauge group $\SO(N)$ and $N_f=N$ quarks in the fundamental representation of the group, plus additional model-dependent singlets, required to avoid unwanted massless particles. The unbroken global flavor symmetry group is of the form
\begin{equation}
G_f=\SO(5)\times H, 
\label{eq:Gf}
\end{equation}
where $H$ is a model-dependent factor.
The SM gauge group is obtained by weakly gauging an $\SU(2)\times \U(1) \subset \SO(5)$ and an $\SU(3)\times \U(1)\subset H$. The superpotential of the theory also includes Yukawa couplings
between the composite sector quarks $Q$ and ordinary SM fields $\xi$, neutral under $\SO(N)$, of the form $\xi Q Q$. 
Such Yukawa couplings, as well as the SM gaugings, explicitly break the flavor symmetry $G_f$.
At low-energies this theory becomes strongly coupled and can be described by a Seiberg dual $\SO(4)$  theory in terms of dual quarks $q$ and mesons $M=QQ$ \cite{Seiberg:1994pq}.\footnote{See \cite{Craig:2011ev} for related ideas in the context of SUSY models with a composite, but not pNGB, Higgs.}
The IR theory has a non-SUSY vacuum where some of the composite quarks develop a Vacuum Expectation Value (VEV), spontaneously breaking $\SO(5)\times \SO(4) \rightarrow \SO(4)_D$  \cite{Intriligator:2006dd}. The Higgs  components are the NGB's along the $\SO(5)/\SO(4)_D$ broken directions.
The superpotential term $\xi QQ$ flows in the IR to a mass mixing term of the form $\xi M$, realizing the partial compositeness scenario.
The $\SO(4)$ gauge fields are identified with the vector mesons present in the bottom-up pCHM.

The spontaneous SUSY breaking in the above vacuum is not enough to give acceptable masses to the SM spartners.
We assume an external source of SUSY breaking  that pushes SM squarks, sleptons and gauginos to large enough masses, and
neglect the contribution to the radiatively generated SM and composite soft terms coming from the spontaneous SUSY breaking.
We do not specify neither the origin of this extra breaking of SUSY nor its mediation, and treat it by adding tree-level soft terms in the UV theory. 
We follow their RG flow from the UV to the IR using the techniques developed in \cite{ArkaniHamed:1998wc} (see also \cite{Abel:2011wv}).  
The soft terms in the composite sector are assumed to be $G_f$ invariant or small, otherwise they would contribute at the leading order to the Higgs mass term and lead to fine-tuning problems.

We construct two models within this context. The first is based on an $\SO(11)$ gauge theory  with global flavor symmetry group $G_f = \SO(5)\times \SO(6)$.
Both the left-handed and right-handed top quark components are elementary and mix with fermion resonances in the composite sector. For simplicity, we assume that the extra source of 
SUSY breaking only affects the visible sector and is transmitted to the composite sector through the SM gauge couplings and the mass mixing parameters.
The latter are effectively of the general form described in \cite{Marzocca:2012zn}. 
The second model is based on an $\SO(9)$ gauge theory with flavor symmetry group $G_f = \SO(5)\times \SU(4)$. The right-handed top quark is fully composite and is identified
with the fermion component of a meson bound state. Soft terms in the composite sector are now necessary to give a reasonable mass to the stop $\widetilde t_R$.

Most qualitative properties of the bottom-up pCHM constructions are derived within our UV models.
Gauge and Yukawa couplings govern the vector and fermion resonance masses, respectively.
A mild splitting between these masses,  favored in bottom-up pCHM constructions  \cite{Redi:2012ha,Marzocca:2012zn} to get a 125 GeV Higgs,  is natural and in general expected.
Contrary to the phenomenological models, our composite sector also contains scalar bound states. 

There are various directions in which our construction can be improved.  It would be nice to have a more complete description of the external source of SUSY breaking and a mechanism that allows 
to have a less SUSY, yet calculable, composite sector.  The main open issue is the appearance of Landau poles at relatively low energies for the SM gauge couplings. 
In analogy to what happens in SUSY models with direct gauge mediation of SUSY breaking, these poles arise from the unavoidable multiplicity of exotic matter fields
coming from the composite sector and charged under the SM groups.
In the two explicit models we construct, they can be pushed at energies higher than the mass of the heaviest particle in the UV completion.
We cannot however consider them as the ``ultimate" UV completions of pCHM (modulo the external SUSY breaking, of course) below the Planck scale. For simplicity we have considered in our paper only the mixing of the top quark with the composite sector.  Extending the construction to all
SM fermions is straightforward, but would significantly exacerbate the Landau pole problem, resulting in unacceptably too low poles for the SM couplings, due to the large number of flavors involved in the composite sector. The Landau poles problem seems quite generic and directly related to the idea of partial compositeness, at least in the case in which one assumes a calculable description of the composite sector in terms of (relatively) weakly coupled resonances.

The structure of the paper is as follows: in section \ref{sec:setup} we introduce the general set-up underlying our models; in section \ref{sec:modelI} we introduce the model I, with a semi-composite $t_R$, estimate the lifetime of the  metastable non-SUSY vacuum in subsection \ref{subsec:VD} and compute the Landau poles for the SM couplings in subsection \ref{subsec:LP}; a similar analysis is repeated in section \ref{sec:ModelII} for the model II, with a fully composite $t_R$; in subsection \ref{sec:VDII} we argue that the metastable vacuum is long-lived by showing the absence of SUSY vacua where it could tunnel to and in subsection \ref{sec:LPII} we compute the Landau poles for the SM gauge couplings;
in section \ref{sec:Pheno} we give a closer look at the connection between the above UV models and the phenomenological pCHM considered in the literature;
we discuss open questions and conclude in section \ref{sec:Conclu}; two appendices complete the paper; in appendix \ref{app:groupGen} we report our conventions for the group generators;
in appendix \ref{app:RGflow} we review the RG flow of soft terms in $\mathcal{N}=1$ SUSY gauge theories, and apply the results to our context.

\section{The Basic Construction}
\label{sec:setup}

The key points underlying our models are best illustrated in a set-up where we keep only the essential structure and remove important, but model-dependent, details.
We focus on constructions where the Higgs is the NGB of an $\SO(5)/\SO(4)$ coset, but the generalization to other cosets should be obvious. 
Consider an ${\cal N}=1$ SUSY $\SO(N)$ gauge theory with $N_f=N$ flavors in the fundamental of $\SO(N)$, with superpotential
\be
W_{el} = m_{ab} Q^a Q^b + \lambda_{IJK} Q^I Q^J \xi^{K}\,.
\label{WGen1}
\ee 
In the first term of eq.(\ref{WGen1}), we split the flavor index $I$ in two sets $I=(i,a)$, $a=1,\ldots,5$, $i=6,\ldots,N$. The fields $\xi^{K}$ are singlets under $\SO(N)$ and in general can
be in some representation of the flavor group $H_f\subset G_f$ left unbroken by the Yukawa couplings $\lambda_{IJK}$. The  $\xi^{K}$'s are eventually identified as the visible chiral fields, such as the
top fields.  We take $\lambda_{IJK}\ll1$, so that these couplings  are marginally relevant, with no Landau poles,  and can be considered as a small perturbation 
in the whole UV range of validity of the theory. We assume the presence of an external source of SUSY breaking, whose origin will not be specified, that produces soft terms for all the SM gauginos and sfermions. 
For simplicity, we neglect for the moment the dynamics of the singlets $\xi^{K}$ and the impact of the external source of SUSY breaking in the composite sector.
We take the quark mass matrix proportional to the identity, $m_{ab}=m_Q \delta_{ab}$, 
to maximize the unbroken anomaly-free global group. For $\lambda_{IJK}=0$, this is equal to 
\be
G_f = \SO(5)\times \SU(N-5)\,.
\label{WGen2}
\ee 
We take $m_Q\ll \Lambda$,  where $\Lambda$  is the dynamically generated scale of the theory.

For $N \leq 3(N-2)/2$, namely $N\geq 6$, the theory flows to an IR-free theory with superpotential \cite{Seiberg:1994pq,Intriligator:1995id}
\be
W_{mag} =  q_I M^{IJ} q_J -  \mu^2 M_{aa} +\epsilon_{IJ K} M^{IJ} \xi^{K}\,,
\label{WGen3}
\ee where 
\be
\epsilon_{IJK} = \lambda_{IJK} \Lambda, \ \ \ \  \mu^2 = - m_Q \Lambda.
\ee
For simplicity, we identify the dynamically generated scales
in the electric and magnetic theories,\footnote{Adopting a notation used in the literature, we often refer to the UV and IR theories as electric and magnetic theories, respectively.}
 whose precise relation is anyhow incalculable. We also set to one the value of the Yukawa coupling of the cubic $qMq$ term in the magnetic theory. 
The fields $q_I$ are the dual magnetic quarks in the fundamental representation of the dual
$\SO(N_f-N+4)_m = \SO(4)_m$ magnetic gauge group, with coupling $g_m$, and $M^{IJ}=Q^I Q^J$ are neutral mesons, normalized to have canonical dimension one. 
The K\"ahler potential for the mesons $M^{IJ}$ and the dual quarks $q_I$ is taken as follows:  
\be
K = {\rm tr}M^\dagger M  + q_I^\dagger e^{V_{mag}}  q_I\,,
\ee
where $V_{mag}$ is the $SO(4)_m$ vector superfield.

The original Yukawa couplings $\lambda_{IJK} Q^I Q^J \xi^K$ in the electric theory flow in the IR to a mixing mass term $\epsilon_{IJK} M^{IJ} \xi^K$ between elementary and composite
fields, the SUSY version of the fermion mixing terms appearing in weakly coupled models with partial compositeness \cite{Contino:2006nn}.
The quark mass term $m_Q Q^a Q^a$, introduced to break the flavor group from $\SU(N)$ down to $\SO(5)\times \SU(N-5)$, is also responsible for a 
spontaneous breaking of supersymmetry by the rank condition, as shown by Intriligator, Seiberg and Shih (ISS) \cite{Intriligator:2006dd}.
Up to global $\SO(5)\times \SO(4)_m$ rotations, the non-supersymmetric, metastable, vacuum is at\footnote{With a common abuse of language, we denote with the same symbol a chiral superfield and its lowest scalar component, since it should be clear from the context the distinction among the two.}
\be
\langle q_m^{n} \rangle  = \mu \, \delta_m^n\,, 
\label{WGen4}
\ee
with all other fields vanishing. For simplicity, in the following we take $\mu$ to be real and positive.
In eq.(\ref{WGen4}) we have decomposed the flavor index $a=(m,5)$, $m,n=1,2,3,4$, and we have explicitly reported the gauge index $n$ as well.
When $\lambda_{IJK}=0$, the vacuum (\ref{WGen4}) spontaneously breaks 
\be
\SO(4)_m\times \SO(5)\rightarrow \SO(4)_D\,,
\label{WGen5}
\ee
where $\SO(4)_D$ is the diagonal subgroup of $\SO(4)_m\times \SO(4)$.
In the global limit $g_m\rightarrow 0$, this symmety breaking pattern results in 10 NGB's:
\bea
{\rm Re}\, (q^m_n - q^n_m)&: & \; {\rm along \; the \; broken} \, \SO(4)_m\times \SO(4) \; {\rm directions} \label{WGen6a}\,, \\
\sqrt{2} \, {\rm Re}\, q_5^n &: & \; \,  {\rm along \; the \; broken} \, \SO(5)/\SO(4)_D \; {\rm directions}\,.
\label{WGen6b}
\eea
For $g_m\neq 0$, the  would-be NGB's (\ref{WGen6a}) are eaten by the $\SO(4)_m$ magnetic gauge fields $\rho_\mu$, that become massive, while the NGB's (\ref{WGen6b}) remain massless
and  are identified with the 4 real components of the Higgs field. 

The remaining spectrum of the magnetic theory around the vacuum (\ref{WGen4}) is easily obtained by noticing that all fields, but
the magnetic quarks $q_5^n$ and the mesons $M_{5n}$, do not feel at tree-level the SUSY breaking induced by the $F$-term of $M_{55}$: 
\be
F_{M_{55}}=- \mu^2.
\label{WGen7}
\ee
The chiral multiplets $(q^m_n+q^n_m)/\sqrt{2}$ and $M_{mn}$ combine and get a mass $2\mu$, as well as 
the multiplets $M_{im}$ and $q_i^m$ that form multiplets with mass  $\sqrt{2}\mu$. 
The chiral multiplets $(q_n^m-q_m^n)/\sqrt{2}$ combine with the $\SO(4)_m$ vector multiplets to give vector multiplets with mass $\sqrt{2} g_m\mu$. 
As we have just seen, the NGB scalar components  ${\rm Re}\, (q^m_n - q^n_m)$ are eaten by the gauge fields, while ${\rm Im}\, (q^m_n - q^n_m)$ get a mass by the $\SO(4)_m$ D-term potential. Similarly, the fermions $(\psi_{q^m_n} - \psi_{q^n_m})/\sqrt{2}$ become massive by mixing with the gauginos $\lambda_{mn}$.
The chiral multiplets $M_{ij}$ and $M_{i5}$ remain massless.

The scalar field $M_{55}$ is massless at tree-level and its VEV is undetermined (pseudo-modulus). This is stabilized at the origin by a one-loop induced Coleman-Weinberg potential, as we will shortly see.
Its fermion partner is also massless, being the Goldstino. Around $M_{55}=0$, the fermions $\psi_{q_5}$ and $\psi_{M_{5m}}$ mix and get a mass $\sqrt{2} \mu$, the scalars $M_{5m}$  get the same mass. 
${\rm Im}\, q_5^m$ get a mass $2\mu$, while ${\rm Re}\,q_5^m$ remain massless, the latter being indeed NGB's. The fate of $M_{55}$ is determined by noticing that the superpotential of the $M_{55}-M_{5m}-q_5^m$ sector is 
\be
W_{mag}\supset -  \mu^2 M_{55} + \sqrt{2}\mu  q_5^m M_{5m} +  (q_5^n)^2 M_{55}\,,
\label{WGen8}
\ee
that is a sum of O'Raifeartaigh models. The associated one-loop potential is well-known (see e.g. appendices A.2 and A.3 of \cite{Intriligator:2006dd}). The pseudo-modulus $M_{55}$ is stabilized at zero, and gets a one-loop  mass
\be
m^2_{M_{55}} = \frac{2(\log 4-1)}{\pi^2} \mu^2\,.
\label{WGen9}
\ee
The SM vector fields are introduced by gauging a subgroup of the flavor symmetry group 
\be
H_f\supseteq \SU(3)_c\times \SU(2)_{0,L}\times \U(1)_{0,Y}
\ee
that is left unbroken when we switch on the couplings $\epsilon_{IJK}$.
We embed $\SU(3)_c$ into $\SU(N-5)$ and $\SU(2)_{0,L}\times \U(1)_{0,Y}$ in $\SO(5)\times \U(1)_X$, where $\U(1)_X$ is a $\U(1)$ factor coming from $\SU(N-5)$ needed to correctly reproduce the SM fermion hypercharges. The details of the embedding are model-dependent and will be considered in the next sections.
We identify $\SU(2)_{0,L}$ as the subgroup of $\SO(4) \cong \SU(2)_{0,L} \times \SU(2)_{0,R}\subset \SO(5)$. 
The hypercharge $Y$ is given by $Y=T_{3R}+X$, where $T_{3R}$ and $X$ are the generators of the $\sigma_3$ direction
$\U(1)_{0,R}\subset \SU(2)_{0,R}$ and of $\U(1)_X$, respectively.  
Denoting by $A_\mu^{aL}$ $(a=1,2,3)$, $A_\mu^{3R}$ and $X_\mu$ the $\SU(2)_{0,L}\times \U(1)_{0,R}\times \U(1)_X$ gauge fields and by $g_0$ (the same for $\SU(2)_L$ and $\U(1)_R$, for simplicity)
and $g_X$ their gauge couplings, we have (see appendix \ref{app:groupGen} for our group-theoretical conventions)
\be
A_\mu^{aL} = W_\mu^a \,, \ \ \ A_\mu^{3R} = c_X B_\mu \,, \ \ \ X_\mu = s_X B_\mu\,,
\label{MixVis1}
\ee
where
\be
	c_X = \frac{g_X}{\sqrt{g_0^2 + g_X^2}}=\frac{g_0^\prime}{g_0}, \quad s_X = \frac{g_0}{\sqrt{g_0^2 + g_X^2}}.
	\label{MixVis2}
\ee
The $\SU(2)_{0,L}\times \U(1)_{0,Y}$ gauge fields $W_\mu^a$ and $B_\mu$ introduced in this way are not yet the actual SM gauge fields, 
because the flavor-color locking given by the VEV (\ref{WGen4}) generates a mixing between the $\SO(4)_m \cong \SU(2)_{m,L} \times \SU(2)_{m,R}$ magnetic gauge fields and the elementary gauge fields.
This explains the subscript $0$ in $\SU(2)_{L,R}$ and $\U(1)_{Y,R}$ and in $g$ and $g^\prime$ in eq.(\ref{MixVis2}).  The combination of fields along the diagonal $\SU(2)_L\times \U(1)_Y \subset \SO(4)_D\times \U(1)_X$ group is finally identified with the SM vector fields. The SM gauge couplings $g$ and $g^\prime$ are given by 
\begin{equation}
\frac{1}{g^2} =  \frac{1}{g_m^2}+\frac{1}{g_0^2} \,, \ \ \ \ \ \ \ 
\frac{1}{g^{\prime 2}} =  \frac{1}{g_m^2}+\frac{1}{g_0^{\prime 2}} \,.
\label{MixVis3}
\end{equation}
This mixing between elementary and composite gauge fields is analogous to the one advocated in bottom-up 4D constructions of composite Higgs models.
The situation is simpler for the color group, since the gauge fields of $\SU(3)_c$ are directly identified with the ordinary gluons of QCD; since the group
 $H$ in eq.(\ref{eq:Gf}) contains $\SU(3)\times \U(1)$, the minimal anomaly-free choices for $H$ are $\SO(6)$ or $\SU(4)$.

The set-up above is still unrealistic because of the presence  of unwanted exotic massless states ($M_{ij}$ and $M_{i5}$).
There are various ways to address these points. We do that 
in the next two sections, where we consider in greater detail the two models  with $H=\SO(6)$ and $H=\SU(4)$, corresponding to $N_f=11$ and $N_f=9$ flavors, respectively.

\section{Model I: a Semi-Composite  \texorpdfstring{$t_R$}{tR}}
\label{sec:modelI}

The first model we consider is based on a SUSY $\SO(11)$ gauge theory with $N_f=N=11$ electric quarks. We also have two additional singlet fields, $S_{ij}$ and $S_{ia}$, transforming
as $({\bf 1},\overline{\bf 20}\oplus {\bf 1})$ and $({\bf 5},\overline{\bf 6})$ of $\SO(5)\times \SU(6)$, respectively.\footnote{See \cite{Green:2010ww} for a similar set-up in the context of models with direct 
gaugino mediation of SUSY breaking.} We add to the superpotential (\ref{WGen1}) the following terms:
\be
 \frac 12 m_{1S} S_{ij}^2+ \lambda_1 Q^i Q^j S_{ij}  + \frac 12 m_{2S} S_{ia}^2+ \lambda_2 Q^i Q^a S_{ia} \,.
\label{Mod1}
\ee
The mass terms in eq.(\ref{Mod1}) break the $\SU(6)$ global symmetry to $\SO(6)$. The total global symmetry of the model is then
\be
G_f = \SO(5) \times \SO(6) \,.
\label{Mod2}
\ee
For $m_{1S,2S}> \Lambda$, the singlets $S_{ij}$ and $S_{ia}$ can be integrated out in the electric theory. 
We get\footnote{Of course, we could have started directly by deforming the superpotential (\ref{WGen1})  with the irrelevant operators quartic in the quark fields
appearing in eq.(\ref{Mod1a}). In the spirit of our paper,  we want to emphasize how easy is to UV complete the above quartic terms.  
See \cite{Kitano:2006xg} for studies of ISS theories deformed by irrelevant operators quartic in the quark fields.} 
\be
W_{el}^{eff} = m_{ab} Q^a Q^b -\frac{\lambda_1^2}{2m_{1S}} (Q^i Q^j)^2-\frac{\lambda_2^2}{2m_{2S}} (Q^i Q^a)^2\,.
\label{Mod1a}
\ee
In the magnetic dual superpotential, the quartic deformations give rise to mass terms for the mesons $M_{ij}$ and $M_{i5}$:
\be
W_{mag} \supset  -\frac 12 m_1 M_{ij}^2 -\frac 12 m_{2} M_{ia}^2 \,,
\label{Mod2a}
\ee
where 
\be
m_i = \frac{\Lambda^2 \lambda_i^2}{m_{iS}}, \ \ \  i=1,2\,.
\label{Mod3a}
\ee 
The mass deformations do not affect the vacuum (\ref{WGen4}), but  obviously change the mass spectrum given in section \ref{sec:setup}.
The multiplets $M_{ij}$  and  $M_{i5}$ are now massive, with masses given by $m_1$ and  $m_2$, respectively, and
the multiplets $M_{im}$ and $q_i^m$  form massive multiplets with squared masses  $(m_{2}^2+16\mu^2 \pm m_{2}\sqrt{m_{2}^2+32\mu^2})/8$. 
We take the masses $m_1$ and $m_2$ as free parameters, although phenomenological considerations 
favour the values of $m_2$ for which the mesons $M_{ia}$, the ones that are going to mix with the elementary SM fields, have a mass around $\mu$.
We summarize in table \ref{tab1} the gauge and flavor quantum numbers of the fields appearing in the electric and magnetic theories.
\begin{table}
\begin{minipage}{8cm}
\begin{center}
\begin{tabular}{|c|c|c|c|} \hline
 &${\SO(11)}_{el}$&${\SO(5)}$&${\SO(6)}$\\ \hline
$Q_i^N$&$\bf{11}$&$\bf{1}$&$\bf{6}$\\
$Q_a^N$&$\bf{11}$&$\bf{5}$&$\bf{1}$\\
$S_{ij}$&$\bf{1}$&$\bf{1}$&${\bf 20}\oplus {\bf 1}$\\
$S_{ia}$&$\bf{1}$&$\bf{5}$&$\bf{6}$ \\ \hline
\end{tabular}
\end{center}
\begin{center}\small{(a)}\end{center}
\end{minipage}
\begin{minipage}{8cm}
\begin{center}
\begin{tabular}{|c|c|c|c|} \hline
 &${\SO(4)}_{mag}$&${\SO(5)}$&${\SO(6)}$\\ \hline
$q_i^n$&$\bf{4}$&$\bf{1}$&$\bf{6}$\\
$q_a^n$&$\bf{4}$&$\bf{5}$&$\bf{1}$\\
$M_{ij}$&$\bf{1}$&$\bf{1}$&${\bf 20}\oplus {\bf 1}$\\
$M_{ia}$&$\bf{1}$&$\bf{5}$&$\bf{6}$\\
$M_{ab}$&$\bf{1}$&${\bf 14}\oplus {\bf 1}$&$\bf{1}$ \\ \hline
\end{tabular}
\end{center}
\begin{center}\small{(b)}\end{center}
\end{minipage}
\caption{Quantum numbers under $G_f$ and the strong gauge group of the matter fields appearing in the composite sector of model I:  (a) UV electric and (b) IR magnetic theories.}
\label{tab1}
\end{table}
We embed $\SU(3)_c$ into $\SO(6)$ and $\SU(2)_{0,L}\times \U(1)_{0,Y}$ in $\SO(5)\times \U(1)_X$, where $\U(1)_X$ is a $\U(1)$ factor coming from $\SO(6)$ (see appendix \ref{app:groupGen}).
We consider in what follows the top quark only, since this is the relevant field coupled to the electroweak symmetry breaking sector.
In terms of the UV theory, we might have Yukawa couplings of the top with the electric quarks, or mixing terms with the singlet fields.  
When the singlets are integrated out, we simply get a shift in the mixing of the top with the meson fields. So, without loss of generality, we can ignore mixing terms 
between the top and the singlets. The most general mixing term is then 
\be
\lambda_L (\xi_L)^{ia} Q_i Q_a + \lambda_R (\xi_R)^{ia} Q_i Q_a \,.
\label{Mix1a}
\ee
We assume in what follows that $\lambda_{L,R}\ll1$ so that the elementary fields do not significantly perturb the above results.   
We have written the mixing terms in a formal $G_f$ invariant way in terms of the fields $\xi_L$ and $\xi_R$. These are spurion superfields, whose only dynamical components are
the SM doublet superfields $Q_L=(t_L,b_L)^t$ and the singlet $t^c$, whose $\theta$-component is the conjugate of the right-handed top $t_R$. In order to write $\xi_L$ and $\xi_R$ in terms of 
$Q_L$ and $t^c$, we have to choose an embedding of $\SU(3)\subset \SO(6)$:
\be
(\xi_L)^{ia} =\left(\begin{array}{ccccc}
b^1 & -i b^1 & t^1 & it^1 & 0 \\
-i b^1 & - b^1 & -it^1 & t^1 & 0 \\
b^2 & -i b^2 & t^2 & it^2 & 0 \\
-i b^2 & - b^2 & -it^2 & t^2 & 0 \\
b^3 & -i b^3 & t^3 & it^3 & 0 \\
-i b^3 & - b^3 & -it^3 & t^3 & 0 \\
\end{array}\right)_{2/3}\!\!, \ \ \ 
(\xi_R)^{ia}  = \left(\begin{array}{ccccc}
0 & 0 & 0 & 0 & (t^c)^1 \\
0 & 0 & 0 & 0 & i(t^c)^1 \\
0 & 0 & 0 & 0 & (t^c)^2 \\
0 & 0 & 0 & 0 & i(t^c)^2 \\
0 & 0 & 0 & 0 & (t^c)^3 \\
0 & 0 & 0 & 0 & i(t^c)^3 \\
\end{array}\right)_{-2/3}\!\!, \ \ \ 
\label{Mix2a}
\ee
in terms of $\SO(6)\times \SO(5)$ multiplets, where the superscript in the fields denote the color $\SU(3)_c$ index. The subscript $\pm 2/3$ denotes the $\U(1)_X$ charge of the fermion.
The terms (\ref{Mix1a}) explicitly break the global group $G_f$ of the composite sector and in the magnetic theory they flow to
\be
\epsilon_L  (\xi_L)^{ia} M_{ia} + \epsilon_R (\xi_R)^{ia} M_{ia} \,.
\label{Mix3a}
\ee
For simplicity, we neglect here the effects induced by possible soft terms present in the electric theory.
We discuss their impact in some detail in appendix \ref{app:RGflow} and in the next section, where we consider a model where they cannot be neglected.
We then add
\be
-{\cal L}_{\cancel{SUSY}} = \widetilde m_L^2 |\widetilde t_L|^2 + \widetilde m_R^2 |\widetilde t_R|^2  +\Big(\epsilon_L B_L (\xi_L)_{ia} M_{ia}+ \epsilon_R B_R (\xi_R)_{ia} M_{ia}  +\frac 12 \widetilde m_{g,\alpha}  \lambda_\alpha \lambda_\alpha+h.c.\Big),
\label{softterms}
\ee
where $\lambda_\alpha$ are the SM gauginos and $\alpha=1,2,3$ runs over the $\U(1)_{0,Y}$, $\SU(2)_{0,L}$ and $\SU(3)_c$ groups.
In order to simplify the expressions below, we take the SM soft terms larger than $\mu$.\footnote{Notice that we cannot 
take the soft terms parametrically large, in particular the stop mass terms, because in this way we would reintroduce a fine-tuning to keep the quadratic Higgs mass term at the electroweak scale.}  
Due to the terms (\ref{Mix3a}) and the interactions with the SM gauginos, the SUSY breaking is transmitted to the composite sector as well.
More in detail, the Dirac fermions $\left(\lambda_{mn},(\psi_{q_n^m}-\psi_{q_m^n})/\sqrt{2}\right)$ mix with the SM gauginos: as a result the former get splitted into two Majorana fermions with masses $\sqrt{2}g_m\mu\pm \delta \widetilde m_{\lambda}$.
Expanding for heavy SM gauginos, we have
\be
\delta \widetilde m_{\lambda,\alpha} \sim \frac{g_\alpha^2 \mu^2}{2 \widetilde m_{g,\alpha}}\,.
\label{maggauginos}
\ee
Similarly, the scalar mesons and magnetic quarks that mix with the stops get soft terms of order
\be
\widetilde m_s^2 \sim - |\epsilon_{L,R}|^2\,,
\label{mstops}
\ee
that tend to decrease their SUSY mass value.
The spectrum of the fields in the $M_{i5}$ and in the $M_{im}$-$q_i^m$ sectors is affected by the
the terms (\ref{Mix3a}), while all the other sectors are unchanged. In the limit of decoupled stops,  we see that a linear combination of fermions given by $t_R$ and the appropriate components of 
$\psi_{M_{ia}}$ remains massless. This field is identified with the actual SM right-handed top. A similar argument applies to $t_L$.
At this stage, the ``Goldstino" $\psi_{M_{55}}$ is still massless. In the case in which we also consider soft terms in the electric $\SO(N)$ theory (see appendix \ref{app:RGflow} for details), the mesons
$M_{ab}$ get a non-vanishing VEV and a mass for $\psi_{M_{55}}$ can be induced from higher dimensional operators in the K\"ahler potential.
Independently of this effect, a linear combination of $\psi_{M_{55}}$ and the Goldstino associated to the external SUSY breaking is eaten by the gravitino, while
the orthogonal combination  gets a mass at least of order of the gravitino mass (see \cite{Cheung:2010mc} for an analysis of Goldstini in presence of multiple sectors of SUSY breaking and specifically
\cite{Craig:2010yf} for a set-up analogous to the one we are advocating here).  We do not further discuss the mechanisms through which $\psi_{M_{55}}$ can get a mass.

\subsection{Vacuum Decay}

\label{subsec:VD}

In presence of the meson mass terms (\ref{Mod2a}),  in addition to the ISS vacuum (\ref{WGen4}), other non-SUSY vacua can appear  \cite{Kitano:2006xg}.
They can be dangerous if less energetic than the ISS vacuum, since the latter can decay through tunneling too quickly to them. 
These vacua do not appear in our model, since the superpotential does not include meson terms of the form $M_{ab}^2$.
Other non-SUSY vacua can be found at $q_{m}^n\sim q_i^n\sim M_{ij}\sim M_{nm}\sim M_{in}\sim\mu$, $q_5^n=0$, $M_{n5}=0$, $M_{i5}=0$, while $M_{55}$ is still a flat direction. They do not lead to the desired pattern of symmetry breaking and they do not allow us to embed the SM in the flavor group. 
All these vacua, even if present,  have however exactly the same tree-level energy of the ISS vacuum and would be irrelevant for the tunneling rate.

Supersymmetric vacua\footnote{By supersymmetric vacua we mean those that are SUSY in the limit where we switch off the external source of SUSY breaking.\label{foot:SUSYsoft}}
are expected when the mesons get a large VEV, in analogy with \cite{Intriligator:2006dd,Kitano:2006xg}.
The scalar potential has a local maximum at the origin in field space, with energy $V_{Max} =5 \mu^4$, while at the local minimum 
$V_{Min}=\mu^4$. We look for SUSY vacua in the region of large meson values, $|M_{ij}|\gg \mu$, $|M_{ab}|\gg \mu$. For simplicity, we take 
\be
 M_{ab} = X\, \delta_{ab}\,, \ \ M_{ij} = Y\, \delta_{ij}\,, \ \ \ \\ M_{ia} = 0\,.
\label{Vac1}
\ee 
For $|X|,|Y|\gg \mu$, the magnetic quarks are all massive and can be integrated out. Below this scale, we get a pure  SUSY $\SO(N)$ Yang-Mills theory with a set of neutral mesons $M$.
The resulting superpotential is 
\be
W =   2 \Lambda^{-\frac 52}({\rm det} M)^{\frac 12}    -  \mu^2 M_{aa}  -\frac 12 m_1 M_{ij}^2 -\frac 12 m_{2} M_{ia}^2 \,, 
\label{Vac2}
\ee
where we neglect  the elementary sector, that gives rise to subleading corrections.
By imposing the vanishing of the $F$-term conditions, we find SUSY vacua at  
\be\begin{split}
X= &  \Lambda^{\frac 56} \mu^{-\frac 13} m_1^{\frac 12} = \epsilon^{-\frac 13}\sqrt{\Lambda m_1} = \epsilon^{-\frac 56}\sqrt{\mu \, m_1} \,,  \\
Y= & \Lambda^{\frac 5{12}} \mu^{\frac 56} m_1^{-\frac 14} = \epsilon^{\frac 56}\Lambda \Big(\frac{\Lambda}{m_1}\Big)^{\frac 14} = \epsilon^{-\frac 5{12}}\mu \Big(\frac{\mu}{m_1}\Big)^{\frac 14}\,,
\label{Vac3}\end{split}
\ee
where
\be
\epsilon=\frac{\mu}{\Lambda}
\label{Vaceps}
\ee
is a parametrically small number.
The vacua (\ref{Vac3}) can also be found directly in the electric theory. In the region where $S_{ij}$ is non-vanishing, all the quarks $Q$ are massive and the theory develops an
Affleck-Dine-Seiberg superpotential of the form \cite{Affleck:1983mk}
\be
W_{np} = (N-N_f-2) \Lambda^{\frac{N_f-3(N-2)}{N_f-N+2}}({\rm det}\, M)^{\frac{1}{N_f-N+2}}\,,
\label{Vac4a}\ee 
where $M=M_{IJ}=Q_I Q_J$. It is straightforward to check that this term induces in fact the SUSY vacua (\ref{Vac3}). The vacuum (\ref{Vac3}) lies in the range of calculability of the magnetic theory  if 
\be
\mu \ll |X|,|Y|\ll \Lambda\,.
\label{Vac5}
\ee
The conditions (\ref{Vac5}), together with the requirement that the mesons $M_{ij}$ are not anomalously light, $m_1\geq \mu$,
determine the allowed range for $m_1$. Parametrizing 
\be
m_1  = \Lambda \epsilon^\kappa\,, 
\label{Vac5a}
\ee
we get
\be
\frac 23 < \kappa \leq  1 \,.
\label{Vac6}
\ee
As a very crude estimate of the lifetime of the metastable vacuum, we can parametrize the potential using the triangular approximation \cite{Duncan:1992ai}, neglecting the direction 
in field space along the $Y$ direction, which is always closer to the ISS vacuum, given the bound (\ref{Vac6}).
The bounce action is parametrically given by \cite{Coleman:1977py,Intriligator:2006dd,Duncan:1992ai}
\be
S_b \sim \frac{|X|^4}{V_{Max}} \sim \epsilon^{-\frac{16}3+2\kappa}\gtrsim \epsilon^{-\frac{10}3}\,.
\label{VacEstimate}
\ee
We conclude that for small $\epsilon$ the metastable vacuum is parametrically long-lived and a mild hierarchy between $\mu$ and $\Lambda$ should be enough to get a vacuum
with a lifetime longer than the age of the universe.

\subsection{Landau Poles} 

\label{subsec:LP}

Similarly to what happens in models with direct gauge mediation of SUSY breaking, where the SM group is obtained by gauging a global subgroup of the hidden sector, one should worry about the possible
presence of Landau poles in the SM couplings, the QCD coupling $\alpha_3$ in particular, due to the proliferation of colored fields.
Our model is no exception and Landau poles develop for the SM gauge couplings $\alpha_i$.  
In order to simplify the RG evolution, we conservatively take all the masses of the magnetic theory to be of order $\mu$, SM superpartners included, 
with the exception of the mesons $M_{ij}$, whose mass $m_1$ is determined in terms of $m_{1S}$ and $\Lambda$. We run from $m_Z$ up to $\mu$ with the SM fields, from $\mu$ up to $\Lambda$ with
the degrees of freedom of the magnetic theory and above $\Lambda$ with the degrees of freedom of the electric theory.

A one-loop computation shows that the $\SU(3)_c$, $\SU(2)_{0,L}$ and $\U(1)_{0,Y}$ couplings develop Landau poles at the scales
\be\begin{split}
\Lambda_3^L = & \, m_{2S}\, \exp\Big(\frac{2\pi}{21\alpha_3(m_Z)}\Big) \Big(\frac{m_Z}{\mu}\Big)^{-\frac{1}{3}}\Big(\frac{\mu}{\Lambda}\Big)^{\frac 27}
\Big(\frac{\Lambda}{m_{2S}}\Big)^{\frac{16}{21}} \,,\\
\Lambda_2^L = & \, m_{2S} \,\exp\Big(\frac{2\pi}{17\alpha_2(m_Z)}\Big)\Big(\frac{m_Z}{\mu}\Big)^{-\frac{19}{102}}
\Big(\frac{\mu}{\Lambda}\Big)^{\frac{22}{17}}
\Big(\frac{\Lambda}{m_{2S}}\Big)^{\frac{11}{17}} \,, \\
\Lambda_1^L = & \, m_{2S} \,\exp\Big(\frac{2\pi}{91\alpha_1(m_Z)}\Big) \Big(\frac{m_Z}{\mu}\Big)^{\frac{41}{546}}
\Big(\frac{\mu}{\Lambda}\Big)^{\frac{336}{546}}
\Big(\frac{\Lambda}{m_{2S}}\Big)^{\frac{215}{273}} \,.
\end{split}\label{LandauPoles}
\ee
We have taken $\lambda_{1,2} \sim 1$ in the superpotential ({\ref{Mod1}), so that $m_{2S}\sim \Lambda/\epsilon$ is the highest scale in the electric theory,
$\alpha_{1,2,3}(m_Z)$ are the $\U(1)_Y\times \SU(2)_L\times \SU(3)_c$ SM couplings evaluated at the $Z$ boson mass $m_Z$.
In deriving eq.(\ref{LandauPoles}) we have matched the $\SU(2)\times \U(1)$ couplings at the scale $\mu$, using eq.(\ref{MixVis3}) with
\begin{equation}
\alpha_m(\mu) = \frac{2\pi}{5\text{log}\left(\frac{\Lambda}{\mu}\right)}\,.
\label{gmagne}
\end{equation}
Notice that the scale of the poles does not depend on $m_{1S}$, since it cancels out  in the contributions coming from $S_{ij}$ and $M_{ij}$. Demanding for consistency that $\Lambda_{i}^L>m_{2S}$
constrains $\epsilon$ to be not too small. This is welcome from a phenomenological point of view, since a too small $\epsilon$ leads to a parametrically weakly coupled magnetic sector (see eq.(\ref{gmagne}))
and too light magnetic vector fields.  On the other hand, $\epsilon$ cannot be too large for the stability of the vacuum, but values 
as high as $1/10$ or so should be fine, given the estimate (\ref{VacEstimate}).  By taking natural choices for $\mu$ around the TeV scale, 
we see that all the Landau poles occur above $m_{2S}$, with $\SU(3)_c$ being the first coupling that blows up, entering the non-perturbative regime 
in the $10^2-10^3$ TeV range.

The Yukawa couplings  $\lambda_{1,2}$ and $\lambda_{L,R}$  in the superpotential (\ref{Mod1}) and (\ref{Mix1a}) might also develop Landau poles.
A simple one-loop computation, in the limit in which the SM gauge couplings are switched off, shows that these poles appear at scales much higher than those defined in eq.(\ref{LandauPoles}). In a large part of the parameter space the Yukawa's actually flow to zero in the UV. 
This is even more so, when the SM gauge couplings are switched on, due to their growth in the UV.

\section{Model II: a Fully Composite \texorpdfstring{$t_R$}{tR}}
\label{sec:ModelII}

The second model we consider is based on a SUSY $\SO(9)$ gauge theory with $N_f=9$ electric quarks and
an additional singlet $S_{ij}$ in the $({\bf 1},{\bf 10})$ of $\SO(5)\times \SU(4)$.
We add to the superpotential (\ref{WGen1}) the following term: 
\be
 \lambda Q^i Q^j S_{ij} \,.
\label{Mod1su4}
\ee
The terms (\ref{Mod1su4}) do not break any global symmetry. The total anomaly-free global symmetry of the model is 
\be
G_f = \SO(5) \times \SU(4) \,.
\label{Mod2su4}
\ee 
In the magnetic theory  eq.(\ref{Mod1su4}) turns into a mass term $\lambda \Lambda M^{ij} S_{ij}$. If we take $\lambda \sim O(1)$ around the scale $\Lambda$,
the singlets $S_{ij}$ and $M^{ij}$ can be integrated out.  At leading order in the heavy mass, this boils down to remove the chiral fields $S_{ij}$ and $M^{ij}$ from the Lagrangian.
We summarize in table \ref{tab2} the gauge and flavor quantum numbers of the fields appearing in the electric and magnetic theories. 

The mass spectrum is the same as given in section \ref{sec:setup}, with the exception of the multiplet $M^{ij}$ that has been decoupled together with the singlet $S_{ij}$.
The multiplet $M_{i5}$ is massless.
We embed $\SU(3)_c\times \U(1)_X$ into $\SU(4)$ and  $\SU(2)_{0,L}\times \U(1)_{0,Y}$ into $\SO(5)\times \U(1)_X$.
The $\U(1)_X$ is identified as the diagonal $\SU(4)$ generator not contained in $\SU(3)_c$, properly normalized, so that ${\bf 4} \rightarrow {\bf 3}_{2/3}\oplus {\bf 1}_{-2}$ under $\SU(3)_c\times \U(1)_X$.
We identify $t_R$ as the (conjugate) fermion component of $M_{\alpha 5}$, $\alpha=6,7,8$.  We also get an unwanted extra fermion, coming from $M_{95}$. Being an $\SU(2)_L$ singlet, $\psi_{M_{95}}$ corresponds to an exotic particle with hypercharge $Y=X=2$. We can get rid of this particle by adding to the visible sector a conjugate chiral field
$\psi^c$ that mixes with $M_{95}$, in the same way as $M_{ia}$ is going to mix with $t_L$. The field $\psi^c$ is actually necessary for the consistency of the model, so that all anomalies cancel.
In the UV theory, the mixing terms are
\be
 \lambda_t \xi^{ia} Q_i Q_a + \lambda_\phi \phi^{ia} Q_i Q_a \,.
\label{Mix1asu4}
\ee
Like in the previous section, we have written the mixing terms in a formal $G_f$ invariant way by means of the superfields $\xi$ and $\phi$. These are spurions, whose only dynamical components are
the SM doublet $Q_L$ and the singlet $\psi^c$. More explicitly, we have
\be
\xi^{\alpha a} =\frac{1}{\sqrt{2}}\left(\begin{array}{c}
b_L \\ -i b_L \\ t_L \\ it_L \\  0 \\
\end{array}\right)_{\!\! 2/3}\!\!, \ \ \  \xi^{9 a} = 0 \,, \ \ \ 
 \phi^{\alpha a} = 0 \,, \ \ \ 
\phi^{9a}  = \left(\begin{array}{c}
0 \\ 0 \\ 0 \\ 0 \\ \psi^c \\
\end{array}\right)_{\!\!-2}\!\!, \ \ \ 
\label{Mix2asu4}
\ee
where we have omitted the color index in $Q$ and $\psi^c$.  In the magnetic theory the Yukawa's (\ref{Mix1asu4}) become
\be
\epsilon_t  \xi^{ia} M_{ia} + \epsilon_\phi \phi^{ia} M_{ia} \,.
\label{Mix3asu4}
\ee
\begin{table}
\begin{minipage}{8cm}
\begin{center}
\begin{tabular}{|c|c|c|c|}\hline 
 &${\SO(9)}_{el}$&${\SO(5)}$&${\SU(4)}$\\ \hline
$Q_i^N$&$\bf{9}$&$\bf{1}$&$\overline{\bf{4}}$\\
$Q_a^N$&$\bf{9}$&$\bf{5}$&$\bf{1}$\\
$S_{ij}$&$\bf{1}$&$\bf{1}$&$\bf{10}$\\ \hline 
\end{tabular}
\end{center}
\begin{center}\small{(a)}\end{center}
\end{minipage}
\begin{minipage}{8cm}
\begin{center}
\begin{tabular}{|c|c|c|c|} \hline
 &${\SO(4)}_{mag}$&${\SO(5)}$&${\SU(4)}$\\ \hline
$q_i^n$&$\bf{4}$&$\bf{1}$&$\bf{4}$\\ 
$q_a^n$&$\bf{4}$&$\bf{5}$&$\bf{1}$\\
$M_{ia}$&$\bf{1}$&$\bf{5}$&$\overline{\bf{4}}$\\
$M_{ab}$&$\bf{1}$&${\bf 14}\oplus {\bf 1}$&$\bf{1}$ \\ \hline
\end{tabular}
\end{center}
\begin{center}\small{(b)}\end{center}
\end{minipage}
\caption{Quantum numbers under $G_f$ and the strong gauge group of the matter fields appearing in the composite sector of model II:  (a) UV electric and (b) IR magnetic theories.}
\label{tab2}
\end{table}
Thanks to the last term in eq.(\ref{Mix3asu4}), the multiplets $M_{95}$ and $\psi^c$ combine and get a mass $\epsilon_\phi/\sqrt{2}$.
The assumption of an external source of SUSY breaking affecting only the visible sector cannot work now, because $t_R$ is a fully composite particle, and would result in an unacceptable
light stop $\widetilde t_R$. We then also add SUSY breaking terms in the composite sector, by assuming that they respect the global symmetry $G_f$. 
In order to have a well-defined UV theory, we introduce positive definite scalar soft terms in the electric theory and analyze their RG flow towards the IR following \cite{ArkaniHamed:1998wc}. 
 See appendix \ref{app:RGflow} for all the details on how this is performed and the approximations underlying the procedure.
Neglecting soft masses for the magnetic gauginos and $B$-terms, the non-SUSY IR Lagrangian reads 
\be\begin{split}
-{\cal L}_{\cancel{SUSY}}\,  = &\,\, \widetilde m_L^2 |\widetilde t_L|^2 + \widetilde m_\psi^2 |\widetilde \psi|^2  + (\epsilon_L B_L (\xi_L)_{ia} M_{ia} +\frac 12 \widetilde m_{g,\alpha}  \lambda_\alpha \lambda_\alpha+h.c.) 
\\ & + \widetilde m_1^2 |M_{ia}|^2 + \widetilde m_2^2 |M_{ab}|^2 + \widetilde m^2_3 |q_{i}|^2 - \widetilde m^2_4 |q_{a}|^2  \,,
\label{softtermsSU4}
\end{split}
\ee
where 
\be\begin{split}
\widetilde m_1^2 \, & = \frac{1}{12}(4+2\omega) \widetilde m^2, \ \ \widetilde m_2^2 = \frac{1}{12}(-8+14\omega) \widetilde m^2, \\  
\widetilde m_3^2 \, & = \frac{1}{12}(-8+5\omega) \widetilde m^2, \ \ \widetilde m_4^2 = \frac{1}{12}(-4+7 \omega) \widetilde m^2
\label{softterms1SU4}
\end{split}
\ee
are the soft mass terms for the scalars in the IR theory, determined in terms of the two $\SO(5)\times \SU(4)$ invariant soft terms in the electric theory, 
$\widetilde m_{1el}^2 Q^{\dagger a} Q^a + \widetilde m_{2el}^2 Q^{\dagger i} Q^i$, with $\widetilde m^2\equiv \widetilde m_{1el}^2$ and 
\be
\omega = \frac{\widetilde m_{2el}^2}{\widetilde m_{1el}^2} \,.
\label{softterms2SU4}
\ee
As can be seen from eq.(\ref{softterms1SU4}), there is no choice of $\omega$ for which all the magnetic soft terms are positive definite. If we take $\omega> 8/5$, 
the first three terms in the second row of eq.(\ref{softtermsSU4}) are positive, while the last one is tachyonic. These tachyons are harmless, since the SUSY scalar potential 
contains quartic terms (both in the $F$ and $D$-term part of the scalar potential) that stabilize them. Negative definite quadratic terms for the $q_a$ are already 
present in the SUSY potential, resulting in fact in the vacuum (\ref{WGen4}). The only effect of the Lagrangian (\ref{softtermsSU4}), at the level of the vacuum, is to change the VEV  (\ref{WGen4}):
\be 
\langle q_m^{n} \rangle  = \delta_m^n \mu\rightarrow  \delta_m^n   \sqrt{\mu^2 + \frac 12 \widetilde m_4^2} \equiv \delta_m^n \widetilde \mu \,.
\label{VEVsu4}
\ee
The mass spectrum is modified by the above soft terms and the new vacuum (\ref{VEVsu4}).
The fermions of the multiplets $(q^m_n+q^n_m)/\sqrt{2}$ and $M_{mn}$ combine and get a mass $2\widetilde \mu$, as well as 
the fermions in $M_{im}$ and $q_i^m$ that get a mass $\sqrt{2}\tilde \mu$. The fermions in $(q_n^m-q_m^n)/\sqrt{2}$ combine with the magnetic gauginos to give fermions with mass $\sqrt{2} g_m\widetilde \mu$. 
The scalar spectrum is more involved. We have
\bea
m^2_{R_S} & = &  4\widetilde \mu^2, \ \ m^2_{I_S} = 4 \mu^2\,, \ \ m^2_{R_A} = 0, \ \ m^2_{I_A} = - 2 \widetilde m_4^2 + 2g_m^2 \widetilde \mu^2, \ \ m^2_{M_{ab}} =  4 \widetilde \mu^2 +  \widetilde m_2^2 ,
\label{scalarmasses}  \\
 m^2_{M_{ia}} &  = &   \widetilde m_1^2+2\widetilde\mu^2 (1-\delta_a^5)\,, \ \ m^2_{M_{5n}} = 2 \widetilde \mu^2+ \widetilde m_1^2\,, \ \ 
m^2_{M_{55}} = \widetilde m_2^2 \,, \ \ m^2_{R_5} = 0 \,, \ \ m^2_{I_5}= 4\mu^2   \,, \ \  \nn
\eea
where $R_{S,A}$, $I_{S,A}$, $R_5$ and $I_5$ denote the canonically normalized fields along the mass eigenvalues directions in field space, defined as
\be\begin{split}
R_S = & \,  {\rm Re}\, (q_m^n+q_n^m), \ \ 
I_S =   {\rm Im}\, (q_m^n+q_n^m),\ \  R_A =  {\rm Re}\, (q_m^n-q_n^m), \ \ I_A = {\rm Im}\, (q_m^n-q_n^m), \\
R_5=&\sqrt{2} \, {\rm Re}\, q_5^n, \ \ I_5= \sqrt{2}\, {\rm Im}\, q_5^n. \ \ 
\label{defmasses}
\end{split}\ee
The 10 massless scalars $R_A$ and $R_5$ are the 10 pNGB's. The former are eaten by the magnetic gauge fields, the latter are identified as the 4 Higgs components.
Notice that the soft terms (\ref{softtermsSU4}) induce a negative mass term for $I_A$, which is compensated by a positive term coming from the $D$-term scalar potential. This state is 
always non-tachyonic for $g_m\geq \sqrt{2}$ and for
\be
\widetilde m_4^2 < \frac{g_m^2 \mu^2}{(1-g_m^2/2)}\,, \ \ {\rm for}\ \ g_m<\sqrt{2}\,.
\ee
Due to the presence of the negative soft mass term for $|q_a|^2$, the global group is spontaneously broken also in the limit $\mu=0$. 
But then the unbroken group $\SO(5)$ is enhanced to $\SU(5)$ and the breaking pattern becomes $\SO(4)_m\times \SU(5) \rightarrow \SO(4)_D$, resulting in a total of $24$ NGB's,
(in the global limit), the additional Goldstones being $I_S$ and $I_5$, as evident from eq.(\ref{scalarmasses}).

The above treatment of soft terms as a perturbation of an underlying SUSY theory makes sense only for soft terms parametrically smaller than $\Lambda$.
Notice that we cannot parametrically decouple the scalars in the composite sector, while keeping the fermions at the scale $\mu$, by taking the soft terms $\widetilde m^2$ in the range
$\mu \ll \widetilde m \ll \Lambda$. This is clear from eq.(\ref{VEVsu4}), since in this limit we decouple the whole massive spectrum in the composite sector. 
In order to keep the compositeness scale around the TeV scale and avoid too light scalars, we take the soft term mass scale around $\mu$.
In addition to that, we still have, like in the model I,  an ``indirect" contribution to the composite soft masses coming from the mixing with the elementary sector, as given by
eqs.(\ref{maggauginos}) and (\ref{mstops}).
A linear combination of fermions given by $t_L$ and the appropriate components of 
$\psi_{M_{im}}$ remains massless and is identified with the SM left-handed top. The ``Goldstino" $\psi_{M_{55}}$ is still massless in these approximations. 
See the considerations made in the last paragraph 
of section \ref{sec:modelI}, that  apply also here, for the possible mechanisms giving a mass to this particle.

\subsection{Vacuum Decay}
\label{sec:VDII}

The non-supersymmetric vacuum we have found can be metastable and supersymmetric (in the sense explained in footnote \ref{foot:SUSYsoft}) vacua might appear,  due to non-perturbative effects in the magnetic theory.
Contrary to the model I in section \ref{sec:modelI}, we have not found SUSY vacua in the regime of validity of the magnetic theory.
The only SUSY vacua we found appear in the electric theory. Assuming $S_{ij}\neq 0$
with maximal rank, all electric quarks are massive and the resulting theory develops the non-perturbative superpotential (\ref{Vac4a}).
Taking the ansatz (\ref{Vac1}) for the gauge-invariant meson directions and $S_{ij}=S_0\delta_{ij}$, we get
\be\begin{split}
F_X & =-5 \Lambda^{-6} X^{\frac 32} Y^2 +m = 0 \,, \\
F_Y & = -4 \Lambda^{-6} X^{\frac 52} Y+\lambda S_0 = 0\,, \\
F_S & = \lambda Y = 0.
\label{Vac4su4}
\end{split}
\ee
The only solution to eq.(\ref{Vac4su4}) is the runaway vacuum 
\be
Y\rightarrow 0, \ \ \  S\propto Y^{-\frac 73} \rightarrow \infty, \ \ \ X\propto Y^{-\frac 43}\rightarrow   \infty \,.
\ee
We have found no other SUSY vacua at finite distance in the moduli space and we then conclude that the metastable vacuum (\ref{WGen4}) is sufficiently long-lived, if not absolutely stable.

\subsection{Landau Poles} 
\label{sec:LPII}

Landau poles at relatively low energies are expected also in this model. 
Within the same approximations made in subsection \ref{subsec:LP}, a one-loop computation shows that the $\SU(3)_c$, $\SU(2)_{0,L}$ and $\U(1)_{0,Y}$ couplings develop Landau poles at the scales
\be\begin{split}
\Lambda_3^L = & \,\Lambda \, \exp\Big(\frac{\pi}{2\alpha_3(m_Z)}\Big)\Big(\frac{m_Z}{\mu}\Big)^{-\frac 74} 
\Big(\frac{\mu}{\Lambda}\Big)^{\frac 14}
\,,\\
\Lambda_2^L = & \, \Lambda \,\exp\Big(\frac{2\pi}{9\alpha_2(m_Z)}\Big) \Big(\frac{m_Z}{\mu}\Big)^{-\frac{19}{54}}
\Big(\frac{\mu}{\Lambda}\Big)^{2} \,, \\
\Lambda_1^L = & \, \Lambda \,\exp\Big(\frac{6\pi}{305\alpha_1(m_Z)}\Big) \Big(\frac{m_Z}{\mu}\Big)^{\frac{41}{610}}
\Big(\frac{\mu}{\Lambda}\Big)^{\frac{236}{305}}
 \,,
\end{split}\label{LandauPolesSU4}
\ee
where we have matched the $\SU(2)\times \U(1)$ couplings at the scale $\mu$, using eq.(\ref{MixVis3}) with
\begin{equation}
\alpha_m(\mu) = \frac{2\pi}{3\text{log}\left(\frac{\Lambda}{\mu}\right)}\,.
\label{gmagneSu4}
\end{equation}
The presence of less flavors and singlet fields in the model II with respect to the model I allows for a significant improvement in the UV behaviour of $\alpha_3$, that now blows up 
at extremely  high energies. However, the different embedding of $\U(1)_X$ in the global group gives rise to several fields with hypercharge $|2|$ that significantly contribute to the
running of $\alpha_1$. As a result, the first coupling to blow up is now  $\alpha_1$.   
For a sensible choice of parameters, e.g. $\mu$ around the TeV scale and $\epsilon \sim 1/10$, we see that $\Lambda_1^L$ is about two orders of magnitude higher than $\Lambda$, around $10^3$ TeV.

\section{Connection with Phenomenological Bottom-up Approaches}

\label{sec:Pheno}

In this section we give a closer look at how the pNGB Higgs interacts with the other fields.
The guideline for 4D bottom-up constructions of pNGB composite Higgs models is given by the Callan-Coleman-Wess-Zumino (CCWZ) construction \cite{Coleman:1969sm} in terms of a chiral Lagrangian parametrizing the pNGB degrees of freedom. For the minimal $\SO(5)\rightarrow \SO(4)$ symmetry breaking pattern the construction has been given in \cite{Giudice:2007fh} and subsequently
generalized in \cite{Contino:2011np,Marzocca:2012zn} to include vector and fermion resonances.
First of all, let us better identify the 10 NGB's $\pi^A$ associated to the symmetry breaking pattern $\SO(5)\times \SO(4)\rightarrow \SO(4)_D$. 
When composite $B$-terms are neglected, the NGB's come entirely from the fields $q_b^n$.  We can parametrize them as\footnote{Notice that it is not naively possible to write 
eq.(\ref{Pheno1}) in terms of superfields, because the NGB's are real fields, while the sigma-model fields such as $U$ in eq.(\ref{Pheno2a}) should be promoted to chiral (and hence complex) superfields. A SUSY formulation is however possible by
complexifying the coset space $G/H$. We will not enter into such construction here (see \cite{Bando:1983ab} for a detailed analysis) because SUSY is anyhow broken in the vacuum (\ref{WGen4}).}
\be
q_b^n = \exp\Big(\frac{i\sqrt{2}}{f} h^{\hat a} T_{\hat a} +\frac i{2f} \pi^a T_ a \Big)_{bc} \widetilde q_c^m  \, \exp\Big(\frac i{2f}  \pi^a T_a\Big)_{mn} \,,
\label{Pheno1}
\ee
where $\widetilde q_c^m$ encode all the non-NGB fields. One can check that the parametrization (\ref{Pheno1}) matches eqs.(\ref{WGen6a}) and (\ref{WGen6b}) at linear order in the field fluctuations. The NGB's decay constant $f$ is fixed by demanding that all the NGB's kinetic terms, coming from $|D_\mu q_a^n|^2$, are canonically normalized. One has
\be
f=\sqrt{2}\mu\,.
\label{Pheno2}
\ee
In order to match our theories with the bottom-up pCHM, it is convenient to take the unitary gauge $\pi^a=0$ and work with an effective $\SO(5)/\SO(4)$ coset parametrized by
\begin{equation}
U=\exp\left(i\frac{\sqrt{2}}{f}h^{\hat{a}}T_{\hat{a}}\right)\,.
\label{Pheno2a}
\end{equation}
In this gauge one has, omitting indices,
\be\begin{split}
iU^t D_\mu q = & \, iU^t \Big(\partial_\mu - i (g_0W_{\mu}^{a}T_{aL}+g^{\prime}_0B_{\mu}T_{3R})\Big) U \widetilde q - g_m  \widetilde q  \rho_\mu^a T^a \\
= & \, (d_\mu^{\hat a} T^{\hat a} + E_\mu^a T^a) \widetilde q - g_m  \widetilde q  \rho_\mu^a T^a\,,
\end{split}
\label{Pheno3}
\ee
where $\rho_\mu^a$ are the magnetic vector mesons,
\be\begin{split}
d_{\mu}^{\hat{a}}= & - \frac{\sqrt{2}}{f}(D_{\mu}h)^{\hat{a}}+\ldots \,, \\
E_{\mu}^{a}= & g_0A_{\mu}^{a}+\frac{i}{f^{2}}(h\stackrel{\leftrightarrow}{D_{\mu}}h)^{a}+\ldots 
\end{split}\label{Pheno4}
\end{equation}
are the CCWZ fields and $A_\mu^a$ are defined in eq.(\ref{MixVis1}). 
Plugging the parametrization (\ref{Pheno3}) into the kinetic term $|D_\mu q_a^n|^2$ and setting $\widetilde q_a^n = \mu \delta_a^n$ gives
\be
|D_\mu q_a^n|^2 \supset  \frac{f^2}4 (d_\mu^{\hat a})^2 + \frac{f^2}{2} (g_m \rho_\mu^a - E_\mu^a)^2 \,.
\label{Pheno6}
\ee
The second term in eq.(\ref{Pheno6}) is responsible for the mixing of SM and magnetic gauge fields. 
We can match the terms (\ref{Pheno6}) with the ones appearing in the bottom-up constructions. In the notations and conventions of  \cite{Marzocca:2012zn}, we have
\be
g_m = g_\rho\,, \ \ \ \ f = f_\rho\,.
\label{Pheno7}
\ee
When the Higgs field gets a VEV, say $\langle h^{\hat 4}\rangle \equiv h \neq 0$, the SM gauge bosons get a mass
\be
m_W = \frac{g f}{2}\sin\frac{\langle h \rangle}{f}\equiv \frac{g v}2\,, \ \ \ \ m_Z = \frac{m_W}{\cos \theta_W} \,,
\label{Pheno8}
\ee 
where $\tan \theta_W = g^\prime/g$, in terms of the canonical SM couplings (\ref{MixVis3}). As expected, the tree-level $\rho$-parameter equals one, thanks to the custodial
symmetry underlying the theory.

Ignoring the SM gauge couplings and the mass mixing in the superpotential, the Higgs can be completely removed from the non-derivative part of the Lagrangian (including the $\SO(4)_m$ $D$-term potential) by a field redefinition of all bosons and fermions with $\SO(5)$ flavor indices: 
\be
M_{ab}\rightarrow (U M U^t)_{ab}, \ \ \ \ \psi_{M_{ab}}\rightarrow (U \psi_{M} U^t)_{ab} \,,
\label{Pheno9}
\ee
and so on.  Notice that complex conjugate fields also transform with the matrix $U$, the latter being real:
$U=U^*$. The Higgs appears in the $\SU(2)_{0,L}\times \U(1)_{0,Y}$ $D$-terms when the SM gauge couplings are turned on. 
The lowest-order interactions involving the Higgs are trilinear couplings
of the schematic form $h \tilde q^2$. In particular, no tree-level Higgs potential can be induced by the scalar interactions in the $D$-term potential.

The field redefinitions like eq.(\ref{Pheno9}) affect the kinetic terms of the fields. Focusing on a specific 2-component fermion, say $\psi_{M_{ia}}$, we get
\be
\psi^\dagger_{M_{ia}} i \bar \sigma^\mu D_\mu \psi_{M_{ia}} \rightarrow 
\psi^\dagger_{M_{ia}}  U^t i \bar \sigma^\mu D_\mu (U \psi_{M_{ia}}) = \psi^\dagger_{M_{ia}} i\bar \sigma^\mu \Big( \nabla_\mu^{ij} \delta_{ab}
-i (d_\mu)_{ab} \delta_{ij} \Big) \psi_{M_{jb}}
\label{Pheno10}
\ee
where 
\be
 \nabla_\mu^{ij} = \delta_{ij}(\partial_\mu - i E_\mu )- i  X_{ij}  g^\prime_0 B_\mu \,.
 \label{Pheno11}
 \ee
Similar considerations apply to the other scalar and fermion fields in the
composite sector. The magnetic quarks would also feature in the covariant derivative the vector mesons $\rho_\mu$.
When the $B$-terms in the composite sector are considered, the mesons $M_{ab}$ develop a VEV, eq.(\ref{RGflow14}). 
The Higgs NGB's come from a combination of the dual quarks $q_a$ and the mesons $M_{ab}$, and correspondingly a
parametrization similar to that in eq.(\ref{Pheno1}) applies to $M_{ab}$ as well. The Higgs kinetic term arises now
from the sum of the $|D_\mu q_a|^2$ and $|D_\mu M_{ab}|^2$ terms. We do not further discuss the deformations induced by the meson VEV's.

After the field redefinitions (\ref{Pheno9}), the fermion mass mixing terms become of the form $\xi U M$ and explicitly depend on the Higgs field. 
In the model I, $\psi_{M_{in}}$ mix with $\psi_{q_i^n}$. The $\bf{6}$ of $\SO(6)$ splits in two fields in the ${\bf 3}$ and ${\bf \bar 3}$ of $\SU(3)_c$, both in the ${\bf 5}$ of $\SO(5)$, that combine pairwise in Dirac mass terms. In total we have two Dirac fermions $Q_i$ in the ${\bf 4}\cong ({\bf 2},{\bf 2})$ of $\SO(4)_D\cong SU(2)_L\times SU(2)_R$, coming from $M_{in}$ and $q_i^n$, and one Dirac fermion singlet $S$, coming from $M_{i5}$. The canonical mass basis requires an $\SO(2)$ rotation among the fields $Q_1$ and $Q_2$: $Q_1 \rightarrow  Q_1 \cos\omega +Q_2 \sin\omega $,  $Q_2 \rightarrow Q_2\cos\omega  -Q_1  \sin\omega$,
where
\be
 \tan \omega = \frac{-m_2+\sqrt{32\mu^2+m_2^2}}{4\sqrt{2}\mu} \,.
\label{Pheno14}
\ee
After this rotation, we see that the fermion mixing is of the general form advocated in \cite{Marzocca:2012zn}, with a mismatch in the number of composite fermion bi-doublets and singlets coupling to the SM fields, $(N_Q=2, N_S=1)$ in the notation of \cite{Marzocca:2012zn}. We can match the mixing (\ref{Mix3a}) with the ones defined in eq.(2.19) of \cite{Marzocca:2012zn}:
\be\begin{split}
\epsilon_{tS} = &\,  \epsilon_R\,, \ \epsilon_{tQ}^1 = \epsilon_R \cos\omega \,, \ \  \epsilon_{tQ}^2 =  \epsilon_R \sin\omega \,, \\
\epsilon_{qS} = &\,  \frac{\epsilon_L}{\sqrt{2}} \,, \ \ \epsilon^1_{qQ} =  \frac{\epsilon_L}{\sqrt{2}} \cos\omega \,, \ \ \epsilon^2_{qQ} =  \frac{\epsilon_L}{\sqrt{2}} \sin\omega\,.
\end{split}\label{Pheno15}
\ee

Fermion mixing in the model II is particularly simple. No diagonalization is needed in the composite sector and  only one (Dirac) fermion bi-doublet couples to $t_L$.
The fields  $t_R$ and $S_L$, and hence the parameters $m_S$, $\epsilon_{tS}$ and $\epsilon_{tQ}$,  should be removed from eq.(2.19) of \cite{Marzocca:2012zn},
being the right-handed top fully composite and identified with $S_R$.
Matching  the remaining mixing gives
\be
\epsilon_{qS}=\epsilon_{qQ} = \epsilon_t\,.
\label{Pheno16}
\ee
See e.g. \cite{Gripaios:2009pe,Marzocca:2012zn} for the mass spectrum after electroweak symmetry breaking and further details.

The Higgs potential, absent at tree level,  is radiatively generated. A double mechanism protects the Higgs mass from quadratic UV corrections, SUSY and its NGB nature.
The models we have constructed resemble two-site models where a collective mechanism further protects the Higgs from quadratic corrections within the IR theory itself.
We notice here that the values of the mixing  (\ref{Pheno15}) and (\ref{Pheno16}) ensure the absence of quadratic divergencies in the matter fermion contribution to the 
radiatively induced one-loop Higgs potential. 
By combining SUSY with the above result,  we see that  the scalar stop + composite contributions to the Higgs potential are free from quadratic divergencies. 
These do not cancel from the gauge contribution, with the value of $f_\rho$ in eq.(\ref{Pheno7}).
We believe that this is due to the fact that quadratic divergencies in the gauge sector would cancel only when adding the contribution coming from the scalars
${\rm Im}\, (q^m_n - q^n_m)$, present in the vector multiplet together with the vector mesons. 
It would be interesting to generalize the Weinberg sum rules discussed in \cite{Marzocca:2012zn} in presence of composite scalars and explicitly verify the above statement.

The cut-off of the magnetic theory is given by
\be
\Lambda = \mu \exp\Big(\frac{2\pi(N_f-6)}{\alpha_m(\mu)}\Big)\,,
\label{Pheno17}
\ee
and can be parametrically higher than $4\pi f$ for sufficiently small magnetic coupling. 
Fermion and vector resonances are governed by different coupling constants.
Roughly speaking 
\be
m_\rho \sim g_m\, \mu, \ \ \ \ \ \ \ m_\psi \sim y \,\mu, 
\ee
where $y$ is the Yukawa coupling of the first term in the superpotential (\ref{WGen3})
(that we have set to one because its actual value is incalculable). This is an interesting property, because the vector resonances are, indirectly by electroweak precision measurements, and directly
by collider searches, constrained to be above the TeV scale. At fixed $\mu$, this  favours not so weak values of $g_m$, in turn giving rise to not so high values of $\Lambda$, see eq.(\ref{Pheno17}).  
On the other hand, a 125 GeV Higgs favours mass scales of the fermion resonances coupled to the top quark to be around or below the TeV scale \cite{Redi:2012ha,Marzocca:2012zn}.

Of course, there is a crucial key difference between our UV completed models and the bottom-up constructions in the literature: given the underlying SUSY, the composite sectors in our models 
include scalar resonances that cannot be decoupled without ruining the calculability in the composite sector.
This can substantially modify the structure of the Higgs potential and the findings of  \cite{Redi:2012ha,Marzocca:2012zn}, as well as other relevant IR properties of pCHM. 
In analogy to the explicit breaking of the chiral symmetry induced by quark mass terms in QCD,
we can also relax the assumption of exactly flavor invariant soft terms in the composite sector, in which case a tree-level Higgs mass term appears.
We do not further discuss the phenomenological consequences of our models, hoping to come back to this important point in a future work.

\section{Conclusions and Open Questions}

\label{sec:Conclu}

We have introduced a framework to provide UV completions of bottom-up composite Higgs models with a pNGB Higgs and partial compositeness.
The set-up is based on Seiberg duality and the existence of (meta-)stable vacua in the IR regime of SUSY gauge theories, where
a spontaneous  breaking of global symmetries occurs. We have presented two models of this kind, their main difference being
the nature of the right-handed top: semi-composite in one model, fully composite in the other. The electroweak SM gauge fields are a mixing of elementary gauge fields
that come from gauging a subgroup of the global group and the gauge fields of the magnetic theory.
The mass mixing between elementary SM and composite fermions have their origin in the UV as trilinear Yukawa couplings between
the elementary SM fields and the electric quarks of the underlying gauge theory.
It is worth to emphasize the simplicity of  our framework, as well as of the two models constructed.

As we have already mentioned in the introduction and in the rest of the paper, there are several theoretical open issues that should be addressed before claiming
of having a complete successful completion of pCHM. It would be nice to have a working model of the extra SUSY breaking that gives SM soft terms and, at the same time,
produces flavor invariant and/or small soft terms in the composite sector. These requests rule out sources of SUSY breaking that are SM gauge mediated to the visible and to the composite sectors.
We might assume one or more hidden sectors where SUSY is broken and is gravitationally transmitted to the visible and the composite sectors. 
However, we would naively have a reincarnation of the SUSY flavor problem (though considerably less severe) in explaining why the 
soft terms in the composite sector are approximately flavor universal or smaller than the ones in the visible sector.
Alternatively, one can assume that the UV theory is the IR description of a yet more fundamental theory where the electric soft terms are suppressed by an RG flow
(as happens at the edge of the conformal window for $N_f = 3/2 N$ in $\SU(N)$, or $N_f = 3/2(N-2)$ in $\SO(N)$, see e.g.\cite{Csaki:2012fh}), although
we are aware that it is not easy to suppress soft terms in this way, see \cite{Buican:2012ec} for a recent analysis.

The most pressing open problem is the occurrence of Landau poles at not so high energies, in the $10^2-10^3$ TeV range for an Higgs compositeness scale $\sim$ TeV.
These poles can be kept above any other mass scale present in the models I and II, but they apparently forbid a naive extension of our set-up to accommodate
all remaining SM fermions. Focusing for simplicity on up-quarks only and the model I, for instance, one might extend the flavor group to be of the form
$G_f = \SO(5)\times H\times H \times H$, where $H=\SO(6)$ and $N=N_f=23$. We gauge an $\SU(3)\times \U(1)_X$ subgroup of $H_D$, 
the diagonal component of $H^3$. All the results presented in section 3  continue to apply, with obvious modifications. Each
fermion component of the mesons $M_{ia}$ mixes with a different up quark, as implied by the partial compositeness paradigm. By further extending the flavor group one can analogously accommodate
fermion resonances that mix with down quarks and leptons. It is clear that the significant proliferation of fields in the composite sector leads to a drastic reduction of the scale where the Landau poles
(\ref{LandauPoles})  occur,  certainly below the scale $\Lambda$ for $\alpha_3$, so that the electric theory is ill-defined.
A possible solution is to give up partial compositeness for the light fermions and assume that they get mass from irrelevant operators of the form
$\epsilon_{ab} \xi_L M_{ab} \xi_R$. These operators in the UV come from quartic superpotential terms of the form $\lambda_{ab} \xi_L Q_{a}Q_b \xi_R$.
When the mesons $M_{ab}$ develop tadpoles, they provide a mass for the SM fermions. Of course, one should now find an alternative
solution to the flavor bounds.
Realizing UV completions of pCHM with all SM fermions partially composite remains an open problem.

The generalization of our results to other cosets or to models featuring different fermion representations should not be too difficult.
It would also be very interesting to study in more detail the phenomenological consequences of our models, including the impact of 
an almost SUSY composite sector on the radiatively induced Higgs potential.

\section*{Acknowledgments}

We thank Matteo Bertolini, Lorenzo Di Pietro, David Marzocca, Flavio Porri and Riccardo Rattazzi for useful discussions.

\appendix

\section{\texorpdfstring{$\SO(6)$, $\SU(3)$ and $\SO(5)$ Generators}{SO(6), SU(3) and SO(5) Generators}}
\label{app:groupGen}

We show here the group theoretical conventions used in the paper.
Let us denote by 
\be
t^{ab}_{ij} = - t^{ba}_{ij} = \frac i2 ( \delta_i^a \delta_j^b -  \delta_i^b \delta_j^a)
\ee
the $n\times n$ anti-symmetric matrices, labeled by $a,b=1,\ldots,n,$ with
matrix elements $i,j$. The matrices $t^{ab}$ have $(+i/2)$ in the $a$-th row and $b$-th column and $(-i/2)$ in the $b$-th row and $a$-th column, with all other components zero.
The $\SO(6)$ generators are taken to be, for $n=6$,
\bea
T^1 & =  & \, t^{32}+t^{14}, \  T^2 = t^{31}+t^{42}, \ T^3 = t^{12}+t^{43}, \   T^4=  t^{16}+t^{52}, \ T^5 = t^{51}+t^{62},\\
 \ \ T^6 & = &\, t^{36}+t^{54}, \   T^7 =  t^{53}+t^{64}, \  T^8 = \frac 1{\sqrt{3}} (t^{12}+t^{34}+2t^{65}), \ T^9 = t^{36}+t^{54}, \ 
 T^{10} =  t^{14}+t^{23},\nn \\
 T^{11} & = & \, t^{24}+t^{31}, \ T^{12} =  t^{16}+t^{25}, \ T^{13} = t^{36}+t^{45}, \ T^{14} = t^{46}+t^{53}, \ T^{15} = \sqrt{\frac 23} (t^{12}+t^{34}+t^{56}). \nn
\eea
In this basis, $T^{1,\ldots,8}$ generate $\SU(3)_c$. The $\U(1)_X$ generator is given by $(4/\sqrt{6})\, T^{15}$, 
so that the fields $\xi_L$ and $\xi_R$ have $\U(1)_X$ charges $2/3$ and $-2/3$, respectively.

The $\SO(5)$ generators are also expressed in terms of the matrices $t^{ab}$ with $n=5$. We take
\be\begin{split}
T^1_L &=  t^{32}+t^{41}, \ \ T^2_L =  t^{13}+t^{42}, \ \ T^3_L = t^{21}+t^{43}, \ \   \\
T^1_R&=  t^{32}+t^{14}, \ \ T^2_R = t^{13}+t^{24}, \ \ T^3_R = t^{21}+t^{34}, \ \   \\
T^{\hat a} & =   \sqrt{2}\, t^{a5}, \ \ \ \ \ \ \hat a = 1,2,3,4 \,.
\label{SO5gendef}
\end{split}\ee
In this basis, $T^{1,2,3}_L$ generate $\SU(2)_L$ and $T^{1,2,3}_R$ generate $\SU(2)_R$ of the $\SO(4)\cong \SU(2)_L\times \SU(2)_R$ local isomorphism. 
The matrices $t^{\hat 1,\hat 2,\hat 3,\hat 4}$ generate the coset $\SO(5)/\SO(4)$. 
A multiplet $\Psi_5$ in the ${\bf 5}$ of $\SO(5)$ decomposes as ${\bf 5} = ({\bf 2,2}) \oplus ({\bf 1,1})$ under $\SU(2)_L\times \SU(2)_R$ and can be written as follows:
\be
\Psi_5 =\frac{1}{\sqrt{2}}\left(
\begin{matrix}
d_- - u_+ \\
- i (u_+ + d_-) \\
u_- + d_+ \\
 i(u_- - d_+) \\
\sqrt{2} s
\end{matrix}\right), \ \
\ee
where 
\be
q_\pm =\left(
\begin{matrix}
u_\pm \\
d_\pm 
\end{matrix}\right) \ \
\label{doubLRDef}
\ee
are the two doublets with $T_{3R} = \pm 1/2$, respectively, forming the bi-doublet, and $s$ is the singlet.

\section{Renormalization Group Flow of Soft Terms}

\label{app:RGflow}

In this appendix we briefly review, following \cite{ArkaniHamed:1998wc}, how to understand the fate of UV soft terms 
in a SUSY gauge theory at strong coupling.\footnote{An alternative derivation of the results of \cite{ArkaniHamed:1998wc} has recently 
been formulated \cite{Abel:2011wv}. We follow the original papers because we have found easier in this way to estimate the corrections coming from superpotential effects, although
a reformulation in terms of the flow of conserved currents and of the would-be conserved $R$-symmetry should be possible.}
For concreteness we focus here on $\SO(N)$ gauge theories with $N-2< N_f\leq 3/2(N-2)$  flavors in the fundamental,
admitting a Seiberg dual IR-free description. This is the case of interest for us, but what follows has clearly a wider applicability.  
More specifically, we want to determine the form of the IR soft terms in the magnetic theory in terms of  the electric ones. 
We first consider the case with no superpotential: $W_{el}=0$. Soft terms can be seen as  the $\theta$-dependent terms of spurion
superfields whose lowest components are the wave-function renormalization of the K\"ahler potential and the (holomorphic) gauge coupling constant.
The Lagrangian renormalized at the scale $E$ is
\be
{\cal L}_{el} = \int d^4\theta \sum_{I=1}^{N_f} Z_I(E) Q_I^\dagger e^{V_{el}} Q_I + \Big( \int d^2 \theta S(E) W^\alpha_{el} W_{el,\alpha} + h.c. \Big)\,,
\label{RGflow1}
\ee 
where  
\be\begin{split}
Z_I(E) = & \, Z_{I}^0(E) \Big(1-\theta^2 B_I(E) - \bar\theta^2 B_I^\dagger(E)-\theta^2\bar\theta^2 (\widetilde m_I^2(E) - |B_I(E)|^2)\Big) \,, \\
S(E) = & \,  \frac{1}{g^2(E)} -\frac{i\Theta}{8\pi^2}+\theta^2 \frac{\widetilde m_\lambda(E)}{g^2(E)}
\label{RGflow2}
\end{split}\ee
are the spurion superfields that encode the $B$-terms $B_I$, non-holomorphic mass terms $\widetilde m_I^2$ and the gaugino mass $\widetilde m_\lambda$.
When there is no superpotential, the $B_I$ terms are irrelevant and can be set to zero. 
The Lagrangian (\ref{RGflow1}) is invariant under a $\U(1)^{N_f}$ symmetry  under which
\be
Q_I\rightarrow e^{A_I}  Q_I \,, \ \ Z_I \rightarrow e^{-A_I-A_I^\dagger}   Z_I \,, \ \ S\rightarrow S - \sum_{I=1}^{N_f} \frac{t_I}{8\pi^2} A_I\,,
\label{RGflow3}
\ee 
where $A_I$ are constant chiral superfields and $t_I$ are the Dynkin indices of the representations of the fields $Q_I$, $t_I=1$ for $\SO(N)$ fundamentals.
In terms of these spurions, one can construct the following RG invariant quantities:
\be
\Lambda_S = E e^{-\frac{8\pi^2 S(E)}{b}} \,, \ \ \hat Z_I = Z_I(E) e^{-\int^{R(E)}\frac{\gamma_I(E)}{\beta(R)}dR} \,.
\label{RGflow4}
\ee
In eq.(\ref{RGflow4}), $b=3(N-2)-N_f$ is the coefficient of the one-loop $\beta$-function $\beta(R)$, $\gamma_I$ are the anomalous dimensions of the fields $Q_I$, and $R(E)$ is defined as $S(E)$ in eq.(\ref{RGflow2}), but in terms of the physical, rather than holomorphic, gauge coupling constant.
In terms of $\Lambda_S$ and $\hat Z_I$, one can further construct a $\U(1)^{N_f}$ and RG invariant superfield:  
\be
 I = \Lambda_S^\dagger \Big( \prod_{I=1}^{N_f} \hat Z_I^{\frac{2t_I}b} \Big) \Lambda_S \,.
\label{RGflow5}
\ee
In the far IR, the dynamics of the system is best described by the magnetic theory, whose degrees of freedom are the mesons $M_{IJ} = Q_I Q_J$, the dual magnetic quarks
$q_I$ and the $\SO(N_f-N+4)$ magnetic vector fields $V_{m}$.  We can use the RG invariants $I$ and $\hat Z_I$ and dimensional analysis to write the lowest dimensional operators in the low-energy Lagrangian:
\be\begin{split}
{\cal L}_{mag} = & \, \int d^4\theta \Big( c_{M_{IJ}} \frac{M_{IJ}^\dagger \hat Z_I\hat Z_J  M_{IJ}}{I} + c_{q_I} q_I^\dagger  e^{V_{mag}}\hat Z_I^{-1} (\prod_J \hat Z_J^{\frac{t_J}b}) q_I \Big)  \\
& \, + \int d^2  \theta \Big( S_{m}(E) W^\alpha_{m} W_{m,\alpha} + \frac{q_I M_{IJ} q_J}{\Lambda_S} \Big)+ h.c. \,,
\label{RGflow6}\end{split}
\ee 
where 
\be
S_{m}(E) =  \,  \frac{1}{g^2_m(E)} -\frac{i\Theta_{m}}{8\pi^2}+\theta^2 \frac{\widetilde m_{m,\lambda}(E)}{g_m^2(E)}
\label{RGflow6a}
\ee
is the magnetic version of the spurion $S$ defined in eq.(\ref{RGflow2}). As shown in \cite{ArkaniHamed:1998wc}, these terms are the leading sources of soft terms provided that $\widetilde m_I\ll \Lambda$, condition that will always be assumed.
The last term in the second row in eq.(\ref{RGflow6}) is the induced superpotential in the magnetic theory. 
Demanding the invariance of $W$ fixes the $\U(1)^{N_f}$ charges of the dual quarks $q_I$ to be $Q_I(q_J) = 1/b - \delta_{IJ}$. These, in turn, fix the $\hat Z$-dependence of the
 K\"ahler potential term of the magnetic quarks. The coefficients $c_{M_{IJ}}$ and $c_{q_I}$ are real superfield spurions, the IR analogues of the wave function renormalization constants $Z_I(E)$.
A relation between IR and UV soft terms is achieved by noticing that in the far UV (IR)  the electric (magnetic) theory is free. 
This implies that for sufficiently high $E$, we can identify $\hat Z_I$ with $Z_I$, neglecting quantum corrections, and identify $m_I^2(E)\equiv  \widetilde m_I^2$ with the physical UV electric soft terms.
Similarly, in the far IR, we can neglect the $\theta^2$ and $\theta^4$ corrections induced by quantum corrections to $c_{M_{IJ}}$ and $c_{q_I}$.
We can then compute the IR soft terms by working out the $\theta^2$ and $\theta^4$ terms in the Lagrangian (\ref{RGflow6}). The physical non-holomorphic soft masses for the mesons and magnetic quarks are 
\be
\widetilde m^2_{M_{IJ}} =  \widetilde m^2_I+ \widetilde m^2_J - \frac{2}{b} \sum_{K=1}^{N_f}  \widetilde m_K^2\,, \ \ \ \ \ \ \ \ 
\widetilde m^2_{q_I} =  -  \widetilde m^2_I + \frac 1b  \sum_{K=1}^{N_f}  \widetilde m_K^2\,.
\label{RGflow7}
\ee
As can be argued from eq.(\ref{RGflow7}), positive definite UV soft terms always flow in the IR to tachyonic soft terms for some mesons and/or magnetic quarks \cite{Cheng:1998xg}.
Indeed, the following sum rule holds:
\be
\sum_{I,J=1}^{N_f} \widetilde m^2_{M_{IJ}} + 2N_f \sum_{I=1}^{N_f} \widetilde m^2_{q_I}  = 0\,.
\label{RGflow7a}
\ee
In our derivation we have tacitly taken the dynamically generated scale in the magnetic theory to coincide with the electric one.
This implies that the same $\Lambda_S$ defined in eq.(\ref{RGflow4}) should be expressed in magnetic variables, namely
\be
\Lambda_S = E e^{-\frac{8\pi^2}{b} S(E)}  = E e^{-\frac{8\pi^2 }{b_{m}}S_{m}(E)}\,,
\label{RGflow8}
\ee 
where $b_{m} = 3 (N_f - N+2)-N_f$. 
Identifying the $\theta^2$ components of eq.(\ref{RGflow8}), we get 
\be
\lim_{E\rightarrow 0 } \frac{\widetilde m_{m,\lambda}(E)}{ b_{m} g_m^2(E)}  = \lim_{E\rightarrow \infty } \frac{\widetilde m_{\lambda}(E)}{b g^2(E)}  \,.
\label{RGflow9}
\ee
Notice that the $\theta^2$ term of $\Lambda_S$ introduces $B$-terms coming from both the $D$- and $F$-components of the magnetic Lagrangian ${\cal L}_{mag}$ that 
precisely cancel each other. This is evident by noticing that the holomorphic rescaling 
\be
M_{IJ} \rightarrow \Lambda_S M_{IJ}
\label{RGflow9a}
\ee
removes $\Lambda_S$  
from the leading order Lagrangian  (\ref{RGflow6}).

Let  us now apply these considerations to our specific set-up. We assume that the electric soft terms do not break the $G_f$ symmetry, so we effectively
have two $\U(1)$ symmetries, rotating the quarks $Q_a$ and $Q_i$, and two different soft terms, $\widetilde m_{1}^2 Q^{\dagger a} Q^a + \widetilde m_{2}^2 Q^{\dagger i} Q^i$.
Applying eq.(\ref{RGflow7}) to the model II with $G_f=\SO(5)\times \SU(4)$, with $b=12$, immediately gives the soft terms reported in eq.(\ref{softterms1SU4}).
Let us see the effect of having $W_{el}\neq 0$. For concreteness, consider the following two terms, 
\be
W_{el} = m Q^a Q^a + \frac 12 \lambda Q_i Q_j S_{ij}\,,
\label{RGflow10}
\ee
that appear in both models I and II. We promote 
$m$ and $\lambda$ to chiral superfield spurions in the spirit of considering an external unspecified SUSY breaking mechanism:
\be
m\rightarrow m(1+\theta^2 B_m) \,, \ \ \ \lambda\rightarrow \lambda (1+\theta^2 A_\lambda)\,.
\label{RGflow11}
\ee
We can still set $B_I=0$ in eq.(\ref{RGflow2}), their effect being a redefinition of the $B_m$ and $A_\lambda$ terms in eq.(\ref{RGflow11}).
We can also reabsorb in $B_m$ and $A_\lambda$ the effect of the field redefinition (\ref{RGflow9a}) that would induce additional $B$-like terms
proportional to the gaugino soft terms.
The above $\U(1)^2$ symmetry is unbroken provided $m$ and $\lambda$ transform as follows:
\be
m\rightarrow e^{-2A_1} m\,, \ \ \ \lambda\rightarrow e^{-2A_2} \lambda\,,
\label{RGflow12}
\ee
with $S_{ij}$ invariant. Two further $\U(1)^2$ and RG-invariants can be constructed starting from $m$ and $\lambda$:
\be
I_m = m^\dagger \hat Z_1^{-2} m\,, \ \ \ \ I_\lambda = \lambda^\dagger \hat Z_2^{-2} \lambda\,.
\label{RGflow13}
\ee
The leading order K\"ahler potential for the mesons and the magnetic quarks is still of the form (\ref{RGflow6}), but now $c_{M_{IJ}}$ and $c_{q_I}$ are
unknown functions of $I_\lambda/(16\pi^2)$ and of $I_m/I$.\footnote{These functions are not completely unrelated, since the combination of K\"ahler terms
associated to conserved global currents should precisely match in the UV and IR theories \cite{Abel:2011wv}. We have not studied this flow in detail, since we anyway
neglect the effects of such corrections.} These corrections are sub-leading provided that $m\ll \Lambda$ and
the effective coupling $\lambda/Z_2$, at some UV scale $E$ where the theory is perturbative, is smaller than $4\pi$. Both conditions can be satisfied in our models. 
In first approximation we can then neglect the superpotential corrections to the RG flow of the soft terms. Of course, even when taking $W_{el}=0$, the relations (\ref{RGflow7}) and (\ref{RGflow9}) 
are only valid in the strict UV and IR limits and with vanishing mixing and SM gauge couplings. 
We have not estimated the corrections coming from relaxing the above approximations, assuming they are sub-leading in eqs.(\ref{RGflow7}) and (\ref{RGflow9}).
It would be interesting to perform a more careful analysis to check the validity of this assumption.

There is an important consequence in having a non-vanishing $W_{el}$.
In the IR, the first term in eq.(\ref{RGflow10}) becomes linear in the mesons $M_{aa}$ and the $B_m$ term induces a tadpole for these fields.
The tadpole changes the vacuum structure of the model. 
Extremizing the whole scalar potential, soft terms included, we get\footnote{We take for simplicity $B_m$ to be real.}
\be
\langle q_m^{n} \rangle  = \,  \widetilde \mu \,  \delta_m^n \,, \ \ \ 
\langle M_{mn} \rangle = -  \,\frac{\mu^2 B_m}{4\widetilde \mu^2+\widetilde m_2^2}\delta_{mn}\,, \ \ \ \langle M_{55} \rangle = -\frac{\mu^2 B_m}{\widetilde m_2^2}\,,
\label{RGflow14}
\ee
where $\widetilde \mu$ is defined as the solution of the following cubic equation in $x\equiv \widetilde \mu^2$:
\be
(x-x_0) (4x+ \widetilde m_2^2)^2+2 \mu^2 B_m^2 = 0\,,
\label{RGflow14a}
\ee
where $x_0\equiv \mu^2+\widetilde m_4^2/2$. For $B_m=0$, we recover eq.(\ref{VEVsu4}). The symmetry breaking pattern is still of the form (\ref{WGen5}) and the qualitative analysis made in the main text continues to be valid.
The 4 Higgs pNGB's are now a combination of ${\rm Re}\,q_5^n$ and ${\rm Re}\,M_{n5}$ and all the mass spectrum of the theory is deformed by the tadpoles. 
For simplicity, we have decided to neglect this effect, assuming a negligibly small (net) $B_m$-term.

As discussed in the main text, in the model I we have assumed negligibly small soft terms in the composite sector.
This is not a necessary assumption and can be relaxed, very much as we do in the model II, where they are needed to 
get a sufficiently heavy stop $\widetilde t_R$. The analysis made below eq.(\ref{softtermsSU4}) would apply with obvious modifications.
However, the presence of soft terms in the composite sector affects the analysis of the vacuum decay pursued in subsection \ref{subsec:VD}}.
We have checked the bound on the soft terms in the composite sector (second row of eq.(\ref{softtermsSU4}), with the addition of the $B$-terms in the composite sector)
above which ${\cal L}_{\cancel{SUSY}}$ can no longer be taken as a perturbation of the SUSY scalar potential in the region of large meson VEV's, eq.(\ref{Vac1}).
In particular, we have verified under what conditions the vacuum displacements from the SUSY values, $\delta X/X$ and $\delta Y/Y$, are much smaller than one. 
Comparable bounds arise from the soft terms $\widetilde m_2$, $\widetilde m_{m,\lambda}$ and $B_m$. We get
\be
|\widetilde m_2| \sim |\widetilde m_{m,\lambda}| \sim |B_m| \ll \epsilon^{\frac 43-\frac\kappa 2} |\mu|\,,
\label{RGflow15}
\ee
where $\kappa$ is defined in eq.(\ref{Vac5a}). Given the bound (\ref{Vac6}) on the allowed values of $\kappa$, we see that the soft terms are constrained to be parametrically smaller than $\mu$.
We have numerically explored also the region of soft terms larger than eq.(\ref{RGflow15}), resulting in shifts $\delta X\gtrsim X$, $\delta Y\gtrsim Y$.
Although it is not possible to draw a definite conclusion from this numerical analysis, we believe that the bound (\ref{RGflow15}) is quite conservative, since the
would-be SUSY vacuum energy becomes greater than the one of the ISS-like vacuum (\ref{VEVsu4}), when the soft terms become comparable to (or larger than) $\mu$.
This is intuitively clear by noticing that the dominant source of energy coming from ${\cal L}_{\cancel{SUSY}}$ is the soft term $\widetilde m_2^2 X^2$ (being $|X|\gg |Y|$) 
and this is positive definite. If this is the case, the ISS-like vacuum would become absolutely stable, provided that other non-SUSY vacua with lower energy do not appear 
elsewhere in field space.


\end{document}